\documentclass[conference]{IEEEtran}

\usepackage{cite}

\usepackage{amsmath}
\usepackage{amssymb}
\usepackage{amsfonts}
\usepackage{algorithmic}
\usepackage{pifont}

\usepackage{graphicx}
\usepackage{textcomp}
\usepackage{xcolor}
\usepackage{tikz}
\usepackage{subcaption}

\usepackage{listings}
\usepackage{tcolorbox}
\tcbuselibrary{listings}

\usepackage{tikz}
\usetikzlibrary{shapes.geometric, arrows, positioning}

\usepackage{tabularx}
\usepackage{booktabs}
\usepackage{tabularray}
\UseTblrLibrary{booktabs}

\newcolumntype{P}[1]{>{\centering\arraybackslash}p{#1}}
\newcolumntype{M}[1]{>{\centering\arraybackslash}m{#1}}

\usepackage{enumitem}
\usepackage{cleveref}

\usepackage[normalem]{ulem}

\definecolor{mygreen}{rgb}{0,0.6,0}
\definecolor{mygray}{rgb}{0.5,0.5,0.5}
\definecolor{mymauve}{rgb}{0.58,0,0.82}

\makeatletter
\newcommand\notsotiny{\@setfontsize\notsotiny\@vipt\@viipt}
\makeatother

\tikzstyle{mitm} = [rectangle, rounded corners, text centered, draw=black, fill=gray!30, text width=1em,font=\scriptsize\ttfamily\bfseries\selectfont]
\tikzstyle{ramps} = [rectangle, rounded corners, text centered, draw=black, fill=red!30,text width=3em, font=\small\ttfamily\bfseries\selectfont]
\tikzstyle{arduino} = [rectangle, rounded corners, text centered, draw=black, fill=blue!30,text width=3.75em, font=\small\ttfamily\bfseries\selectfont]

\tikzstyle{arrow} = [thick,->,>=stealth]

\newcommand{\newt}[2]{#2}

\IEEEoverridecommandlockouts

\def\BibTeX{{\rm B\kern-.05em{\sc i\kern-.025em b}\kern-.08em
    T\kern-.1667em\lower.7ex\hbox{E}\kern-.125emX}}

\begin{document}

\title{\textsc{OffRAMPS}: An FPGA-based Intermediary \\ for Analysis and Modification of  \\ Additive Manufacturing Control Systems} 



\author{%
\IEEEauthorblockN{Jason Blocklove\IEEEauthorrefmark{1}, Md Raz\IEEEauthorrefmark{1}, Prithwish Basu Roy\IEEEauthorrefmark{2}, Hammond Pearce\IEEEauthorrefmark{3},\\ Prashanth Krishnamurthy\IEEEauthorrefmark{1}, Farshad Khorrami\IEEEauthorrefmark{1}, Ramesh Karri\IEEEauthorrefmark{1}}
\IEEEauthorblockA{%
\IEEEauthorrefmark{1}\textit{New York University, New York, USA}\\
\IEEEauthorrefmark{2}\textit{NYU Abu Dhabi, Abu Dhabi, UAE}\\
\IEEEauthorrefmark{3}\textit{University of New South Wales, Sydney, Australia}\\
jason.blocklove@nyu.edu, md.raz@nyu.edu, pb2718@nyu.edu, hammond.pearce@unsw.edu.au,\\ prashanth.krishnamurthy@nyu.edu, khorrami@nyu.edu, rkarri@nyu.edu
}
}

\maketitle


\begin{abstract}
Cybersecurity threats in Additive Manufacturing (AM) are an increasing concern as AM adoption continues to grow. AM is now being used for parts in the aerospace, transportation, and medical domains. Threat vectors which allow for part compromise are particularly concerning, as any failure in these domains would have life-threatening consequences. 
A major challenge to investigation of AM part-compromises comes from the difficulty in evaluating and benchmarking both identified threat vectors as well as methods for detecting adversarial actions. In this work, we introduce a generalized platform for systematic analysis of attacks against and defenses for 3D printers. Our ``\textsc{OffRAMPS}'' platform is based on the open-source 3D printer control board ``RAMPS.''
\textsc{OffRAMPS} allows analysis, recording, and modification of all control signals and I/O for a 3D printer.
We show the efficacy of \textsc{OffRAMPS} by presenting a series of case studies based on several Trojans, including ones identified in the literature, and show that \textsc{OffRAMPS} can both emulate and detect these attacks, i.e., it can both change and detect arbitrary changes to the g-code print commands.
\end{abstract}


\section{Introduction}

\begin{figure}[b]
    \centering
    \vspace{-3mm}
    \includegraphics{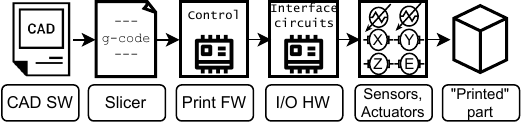}
    \caption{Simplified Additive Manufacturing (3D printing) process. Malicious interference can occur at any step.}
    \vspace{-1mm}
    \label{fig:printing-process}
\end{figure}

Additive Manufacturing (AM), also known as 3D printing, is the process of building up a manufactured component by repeatedly adding  material in specific quantities and locations. Subtractive manufacturing, instead, removes raw material until a final part is left.
AM is performed by designing a part using a computer-aided design (CAD) tool such as Autodesk Fusion or Solidworks, then sending the part to a ``slicer'' program to separate the part into component layers and, based on the target 3D printer, exports \texttt{g-code} which encodes the print head movements used to create the part (see Figure~\ref{fig:printing-process}).

Declining prices  and  increasing quality of modern 3D printers is making them a common tool in hobbyist and professional spaces. In one form of 3D printing, fused filament fabrication (FFF)---also known as fused deposition modeling (FDM) printing---molten plastic is extruded in layers to build the part.
Other forms of 3D printing include stereolithography (SLA) which is the process of shining a UV light in specific shapes per-layer through UV-curing resin to build a part up and selective laser sintering (SLS) which scans a laser  a bed of reactive powder to fuse particles into solid layers. 

The ubiquity and ease of use of 3D printing has grown its adoption within a variety of safety-critical industry sectors, including in the biomedical domain~\cite{placone_recent_2018}, in robotic components~\cite{tawk_design_2021}, construction~\cite{lakhal_robotized_2020}, aerospace~\cite{blakey-milner_metal_2021}, the automotive sector~\cite{vasco_chapter_2021}, and others.
The varying needs of each industry has prompted both innovation and proliferation of interrelated tools and products, from hobbyist to professional and commercial domains. 
%
With this growth in usage comes the growth in \textit{threats} against additive manufacturing systems\newt{}{~\cite{zeltmann_manufacturing_2016,mahesh_survey_2020}}.

The threat landscape in the 3D printing domain encompasses the hardware, software, and supply chain, and can impact every part of the process in Figure~\ref{fig:printing-process}.
Hardware attacks and defenses, which both affect and utilize the integrated circuitry and printed circuit boards used for AM systems, allow for modifications which can internally affect the quality, appearance, or structural integrity of the desired model.
Software-based modifications aim to affect the generated geometry of the design, usually resulting in the export of compromised \newt{control commands, known as}{}\texttt{g-code}.
Finally, supply-chain attacks affecting 3D printers  may include compromised slicing programs, defective components, or inherently flawed filaments and raw materials. 
Unfortunately, the cybersecurity threat landscape in 3D printing remains under-explored, in part due to the complexity involved in studying attacks which impact the real-world physical aspects of additive manufacturing.

\underline{Contributions:}
To address these shortcomings, we introduce \textsc{OffRAMPS}, an FPGA-based integrated test-bed which facilitates the analysis and modification of key signals required to drive the functional components of modern fused filament fabrication (FFF) 3D printers.
It leverages the RAMPS open-source 3D printer control platform, which is representative of commercial offerings while allowing device modification.

\textsc{OffRAMPS} is the \textbf{first platform to support \newt{emulation}{in-hardware analysis} of both attacks and defenses}. 
\textsc{OffRAMPS} enables in-depth exploration of novel printer attack strategies, useful for the identification of previously unexamined security blindspots, and \textbf{we provide a suite of representative Trojans} for this purpose. 
The analysis capabilities allow for expanding the defensive state of the art, and \textbf{we present one defense capable of identifying major Trojans from the literature}.
\textsc{OffRAMPS} is open-source, available here:\cite{zenodo_repo}.
\section{Prior Work}

\subsection{3D Printing Security Threats}

Over the last decade, improvement in the quality of printing materials and printers abetted printing intricate components with ease.
AM found use in diverse fields like aerospace engineering, construction engineering, bio-medical engineering, etc.
With its newly gained popularity, AM has become the target of attackers with malicious intents of sabotage and espionage~\cite{gupta_additive_2020,yampolskiy_security_2018}.
In sabotage-motivated attacks, the attackers either aim to compromise the overall printing process or focus on compromising the quality of the printed product such that its longevity is significantly reduced, causing irreparable damage to the victim company's goodwill.
In their work ``dr0wned''~\cite{belikovetsky_dr0wned_2017}, the authors demonstrated an end-to-end cyber-physical attack that was initiated by introducing malware in the victim's machine.
This malware finds design files in the system, identifies spots that are vulnerable to stress, and inserts sub-millimeter holes in them.
As a demonstration, they compromised the design of a quadcopter drone's propeller, which caused the drone to crash mid-flight.
In another work~\cite{moore_implications_2017}, the authors have modified the Marlin firmware\newt{}{~\cite{marlinfirmware_marlin_2023}} to introduce changes ranging from minor modifications of the executing \texttt{g-code} to the execution of alternative \texttt{g-code}, leading to printing malformed or totally incorrect objects.
The infrequently updated firmware's bootloader becomes an attractive target for stealthy Trojan insertion in the recent study Flaw3D~\cite{pearce2022flaw3d}.
Authors injected Trojans into the bootloader's flash memory, undermining the quantity of extruded material and compromising the print's quality.

In a different class of attack, the attacker aims to exfiltrate information about Intellectual Properties (IPs) being printed.
A 3D printer has mechanical components, such as motors and heating elements that require specific signals to actuate.
A firmware like Marlin, that resides in the controller of the printer, is responsible for parsing the \texttt{g-code} and generating these signals.
The attackers have prior information about the type of motors and can analyze these signals and the power consumed~\cite{gatlin2021encryption} by the motors to gain insight into the linear movements in different axes, thus partially recovering the executing \texttt{g-code}.
In the papers~\cite{al2016acoustic,kcadacousticdet,belikovetsky2019digital}, the authors have exploited a similar correlation between the executing \texttt{g-code} and the sound emanated by the motors and the actuators to partially reverse engineer the IP.
\newt{}{Attackers can also leverage optical side-channels to recreate \texttt{g-code} for a design being printed~\cite{liang2022hiding}}.
These findings highlight risks and security concerns in the 3D printing process.   

\subsection{3D Printing Threat Detection Techniques} A significant number of the currently available threat detection methods are based on side-channel analysis.
Side-channels are passive mediums that leak information about the printing process due to the operation of various physical components of the printing device.
In the case of a 3D printer, the sound emitted by the rotation of the motors, the change in the magnetic field causing the rotation of motors, the power consumed by the motors, the temperature of the hotend, and video of the overall printing process can be considered as the source of acoustic, magnetic, power, thermal~\cite{al2016forensics}, and optical side-channel leakage, respectively.
The assumption for this type of detection is that a good print will have a different side-channel leakage profile than a compromised print. 
In~\cite{kcadacousticdet,belikovetsky2019digital}, acoustic emission of the motors in a secure setup is used as the golden model and is compared to the acoustic signature of future prints to detect anomalies due to \texttt{g-code} manipulation.
However, acoustic side-channels have limited accuracy in determining small and rapid moves, which reduces their efficacy.
In~\cite{gatlin2019detecting} the power profiles of the four stepper motors for the X, Y, and Z axes plus the extruder are used as a golden reference model against the power profile of future prints.
Power side-channel-based analysis, although effective, requires forty repetitions of each print to nullify the noise and syncing issues, making it expensive and non-scalable.
In\newt{}{~\cite{wu_detecting_2019}} video feeds (optical side-channel) of numerous printing processes are used to train an ML model and this  model is used to classify the current print as actual or compromised.
The optical side-channels allow visual reconstruction of the  object and thus can detect layer-by-layer anomalies of the printing process as long as it has been trained to capture such errors.
\newt{}{While quite effective, this method requires significant hardware overhead with very specific requirements for camera placement and image capture.
The accuracy of the results is also dependent on the specific features being printed and the machine learning methodology used to detect the anomalies.}

AM security can be thought of as a subset of cyber-physical systems security. For surveys of this area, see~\cite{dibaji_systems_2019,humayed_cyber-physical_2017}.
\section{OffRAMPS Board Overview}
\begin{figure}
    \centering
    \begin{subfigure}{0.48\linewidth}
        \begin{tikzpicture}
            \node[anchor=south west, inner sep=0] (image) at (0,0) {\includegraphics[width=\linewidth]{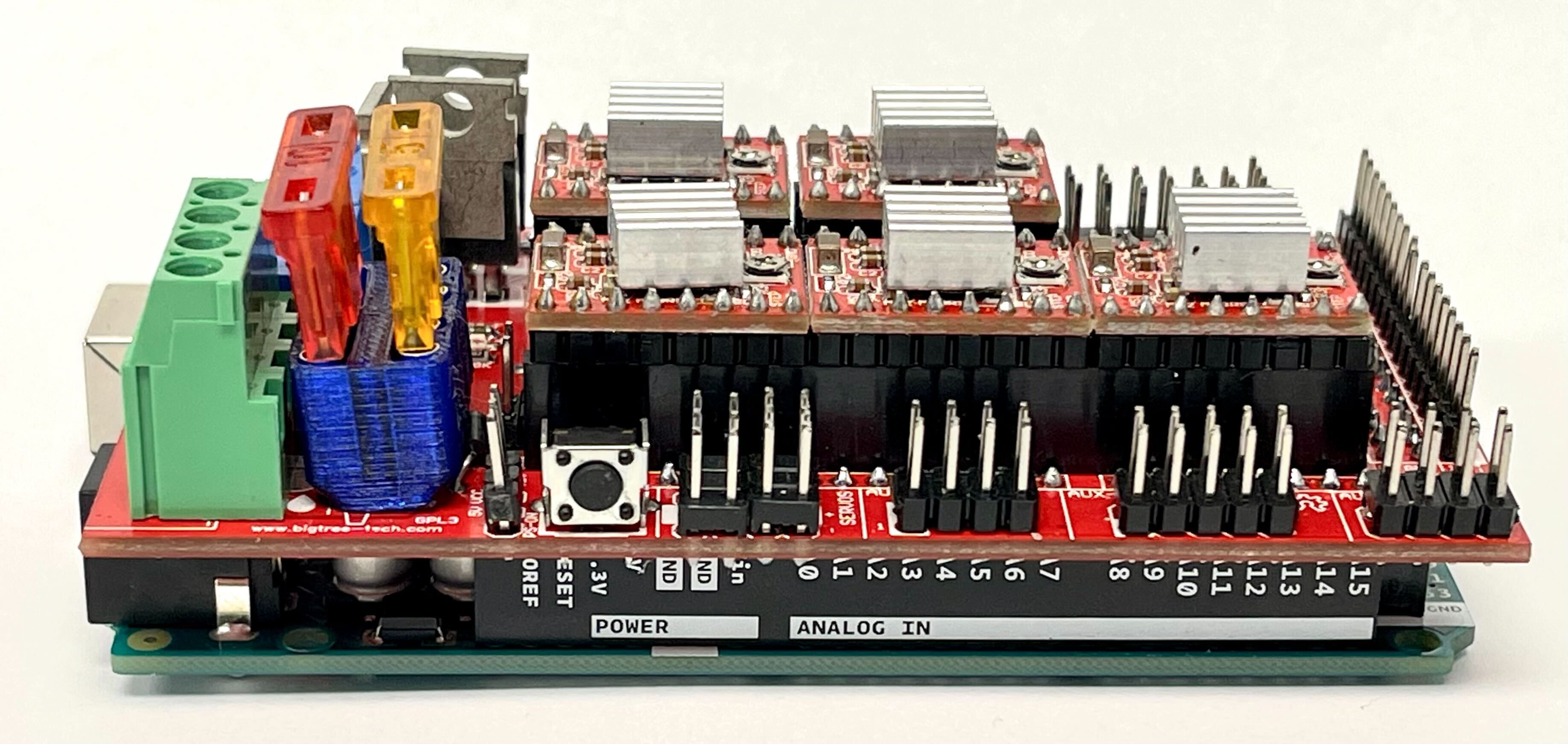}};           
            \begin{scope}[x={(image.south east)},y={(image.north west)}]
                \draw[blue, line width=1.5pt] (0.04,0.05) rectangle (0.95,0.26);
                \draw[red, line width=1.5pt] (0.02,0.28) rectangle (0.98,0.94);
            \end{scope}
        \end{tikzpicture}
        \caption{The standard stack of Arduino Mega (blue box) with RAMPS 1.4 (red box) as a HAT.}
        \vspace{2mm}
        \label{fig:ramps-1}
    \end{subfigure}
    \hspace{1mm}
    \begin{subfigure}{0.48\linewidth}
        \begin{tikzpicture}
            \node[anchor=south west, inner sep=0] (image) at (0,0) {\includegraphics[width=\linewidth]{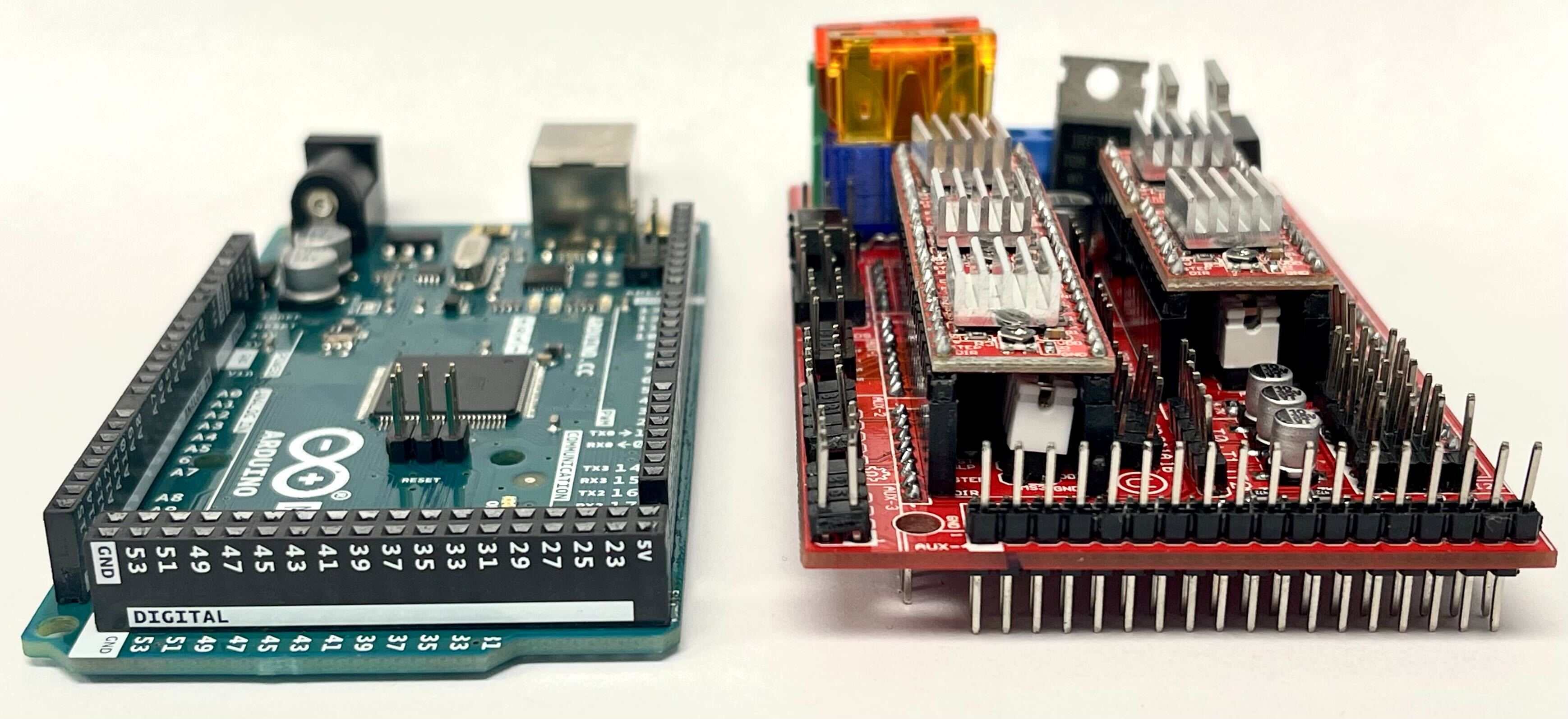}};
            
            \begin{scope}[x={(image.south east)},y={(image.north west)}]
                \draw[blue, line width=1.5pt] (0,0.05) rectangle (0.49,0.97);
                \draw[red, line width=1.5pt] (0.51,0.05) rectangle (1,0.97);
            \end{scope}
        \end{tikzpicture}
        \caption{The Arduino Mega (blue box) and RAMPS (red box) separated to be placed on the \textsc{OffRAMPS}.}
        \label{fig:ramps-2}
        \vspace{2mm}
    \end{subfigure}
    \begin{subfigure}{\linewidth}
        \begin{tikzpicture}
            \node[anchor=south west, inner sep=0] (image) at (0,0) {\includegraphics[width=\linewidth]{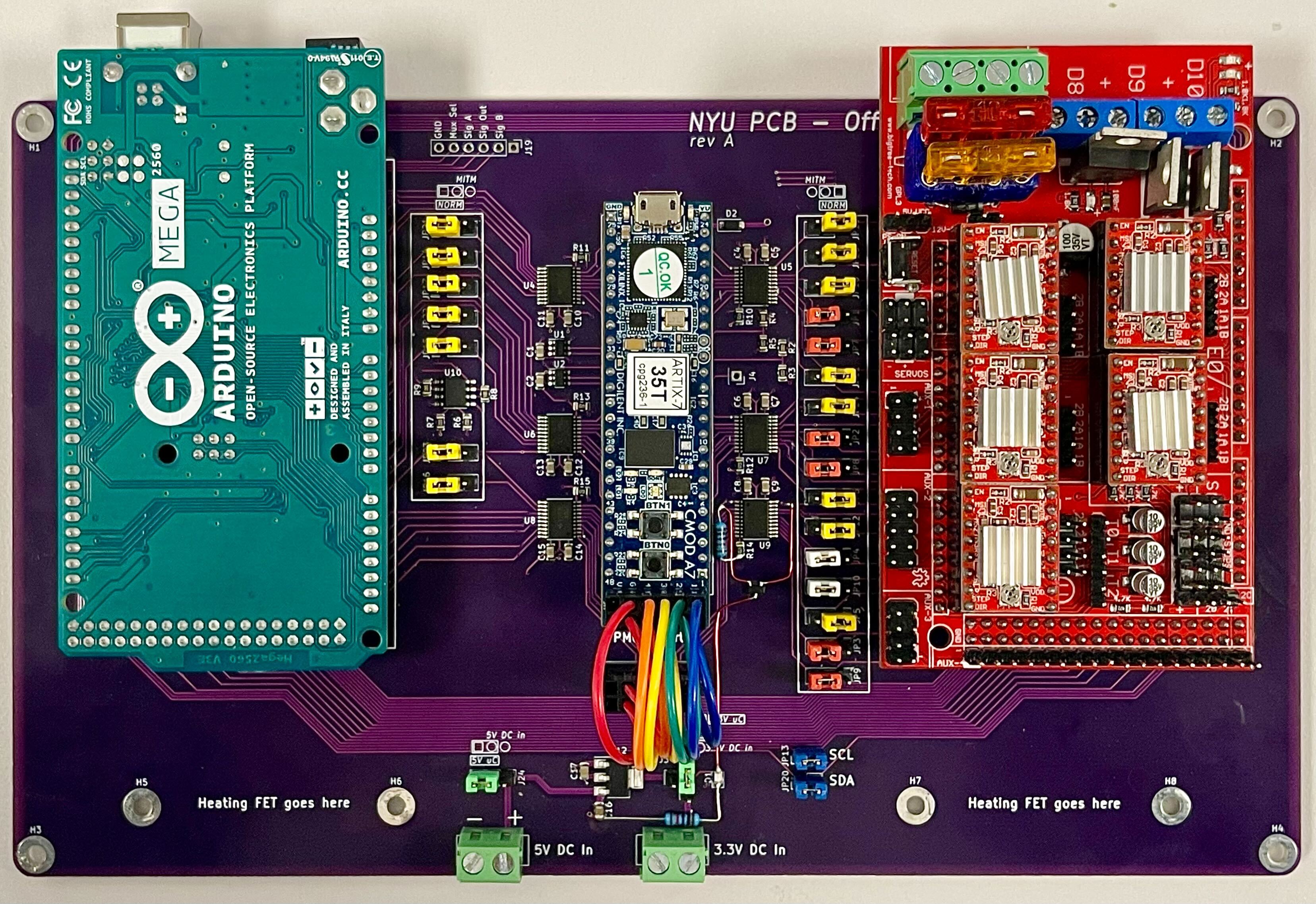}};
            
            \begin{scope}[x={(image.south east)},y={(image.north west)}]
                \draw[blue, line width=2pt] (0.04,0.23) rectangle (0.3,0.98);
                \draw[yellow, line width=2pt] (0.31,0.42) rectangle (0.37,0.8);
                \draw[white, line width=2pt] (0.4,0.16) rectangle (0.6,0.8);
                \draw[yellow, line width=2pt] (0.61,0.24) rectangle (0.66,0.8);
                \draw[red, line width=2pt] (0.67,0.26) rectangle (0.96,0.96);
                \draw[green, line width=2pt] (0.34,0.02) rectangle (0.6,0.15);

                \node[rectangle, draw, blue, fill=white] at (0.175,0.13) {\small\ttfamily\bfseries Arduino Mega};
                \node[rectangle, draw, red, fill=white] at (0.82,0.13) {\small\ttfamily\bfseries RAMPS 1.4};
                \node[rectangle, draw, fill=white, text width=4em, align=center, font=\small\ttfamily\bfseries\linespread{0.8}\selectfont] at (0.5,0.88) {Cmod-A7 FPGA};
                
            \end{scope}
        \end{tikzpicture}
        \caption{\textsc{OffRAMPS} board fully populated. The Arduino Mega (blue box, far left) is flipped upside down to plug into the headers and interfaces with a host computer over USB. The RAMPS 1.4 (red box, far right) receives control signals from the Arduino and sends back certain feedback information. The jumpers (two yellow boxes) determine whether each signal will be passed through the Digilent Cmod-A7 (white box, center) or come directly from the intended source. The power circuitry (green box) allows for Arduino and FPGA power to be derived from several sources as needed.}
    \end{subfigure}
    \caption{The stack of Arduino Mega and RAMPS board separated and put in place on the \textsc{OffRAMPS} board.}
    \label{fig:offramps}
\end{figure}
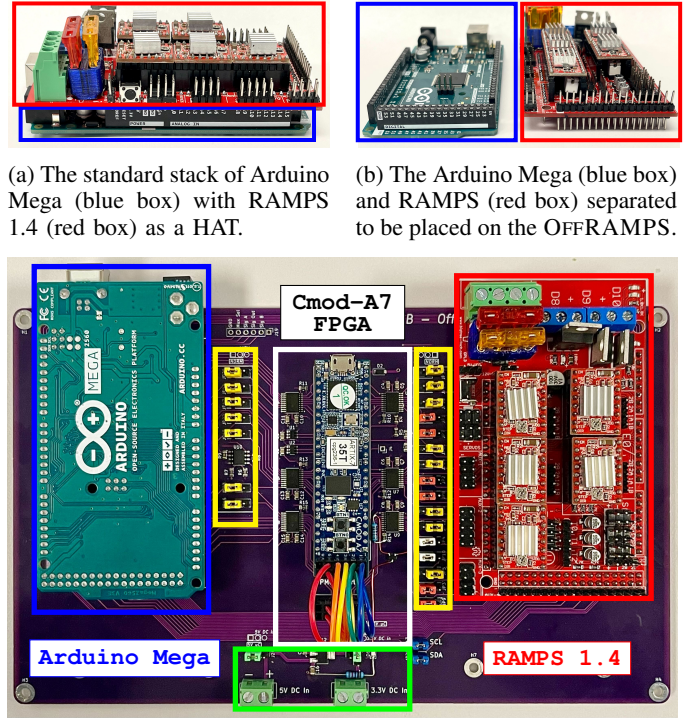

This section details the design of the \textsc{OffRAMPS} board (\Cref{fig:offramps}), a \newt{PCB}{printed circuit board (PCB)} which uses a field programmable gate array (FPGA) as a machine-in-the-middle (MITM) in a popular open-source 3D printer control system.
The board is designed to interface with 1) a Digilent Cmod-A7 FPGA development board, 2) an Arduino Mega running Marlin firmware, and 3) a RAMPS 1.4 3D printer control board.
We add jumpers and logic level shifters to allow the signals to be rerouted as necessary for different experiments.
In Figure~\ref{fig:printing-process}, \textsc{OffRAMPS} is located between the Controller (Print Firmware) and the Interface circuits (I/O Hardware), which enables it to detect and interfere with all signals at digital level voltages (lower than 5V).

\subsection{Open Source 3D Printing}
In the FFF printing space there are numerous companies which make consumer 3D printers such as Creality, Bambu, Prusa, and Ultimaker; however, open-source printers have been a mainstay in 3D printing and many manufacturers (such as Prusa) continue this tradition by open-sourcing their own designs. 
Many of the open-source design components for printer hardware, mechanical and electrical, fall under the RepRap project~\cite{reprap_ramps_2022}. 
The low-cost RepRap Arduino Mega Pololu Shield (RAMPS) printer control board is one such component\newt{, so named because the original version used an Arduino Mega (ATMEGA2560 Chipset) and Pololu stepper driver modules (A4988 Chipset)}{}.
In this work we use version 1.4 of this board, 
which is designed to interface with an Arduino Mega as a hardware-on-top (HAT) device.
The Arduino must run firmware which can interface with a host computer, and for this purpose Marlin is often used as it is another fully open-source piece of the system.
Most designs of RepRap FFF printers can make use of this stack of control boards, making it a prime candidate for evaluation and representative function of other boards.


The broader RepRap project is an endeavor aimed at creating self-replicating 3D printers that are open-source, and has been a major driver for the increased access of additive manufacturing. 
This increased access has also come at a cost, however---clones of these boards are sold by an extensive list of sellers through various online outlets. These clones, which may come with different ICs and functionalities, are of varying quality and provenance. 
Unaware end-users may be impacted negatively by faulty control boards, which may be caused by inferior counterfeit components with undersirable changes to the originals. 

\subsection{\textsc{OffRAMPS}: Design Motivation}
A common threat model for AM comes from malicious third parties modifying the controller \newt{printed circuit boards (PCBs)}{PCBs}~\cite{pearce2022flaw3d,beckwith_needle_2022,moore_implications_2017} to feature firmware or hardware Trojans. Other attacks come from the software space, where CAD programs may be compromised~\cite{belikovetsky_dr0wned_2017}.
It is thus desirable to support emulation of hardware, firmware, and software Trojans in a single platform, as well as provide capabilities to analyze signals passing from the firmware \newt{running}{}on the control CPU to the driving components. 

\begin{figure}
    \centering
    \begin{subfigure}{0.46\linewidth}
        \centering
        \begin{tikzpicture}[node distance=5em]
            \node (arduino) [arduino] {Arduino Mega};
            \node (ramps) [ramps,right of=arduino] {RAMPS 1.4};
        
            \draw[arrow] (arduino.15) -- (ramps.west|-arduino.15);
            \draw[arrow] (ramps.west|-arduino.-15) -- (arduino.-15);
        \end{tikzpicture}
        \caption{Unmodified signal chain.}
        \vspace{0.5em}
    \end{subfigure}
    \begin{subfigure}{0.52\linewidth}
        \centering
        \begin{tikzpicture}[node distance=4.5em]
            \node (arduino) [arduino] {Arduino Mega};
            \node (fpga) [mitm, right of=arduino, minimum height=2em, text width=2.5em] {FPGA Trojan};
            \node (ramps) [ramps,right of=fpga, xshift=-0.375em] {RAMPS 1.4};

            \draw[arrow] (arduino.15) -- (fpga.west|-arduino.15);
            \draw[arrow] (fpga.west|-arduino.-15) -- (arduino.-15);
            \draw[arrow] (fpga.15) -- (ramps.west|-fpga.15);
            \draw[arrow] (ramps.west|-fpga.-15) -- (fpga.-15);
        \end{tikzpicture}
        \caption{FPGA for signal modification.}
        \vspace{0.5em}
    \end{subfigure}
    \begin{subfigure}{0.49\linewidth}
        \centering
        \begin{tikzpicture}[node distance=4.2em]
            \node (arduino) [arduino] {Arduino Mega};
            \node (fpga) [mitm, right of=arduino, text width=3em, yshift=-2em] {FPGA Pulse Capture};
            \node (ramps) [ramps,right of=fpga, yshift=2em, xshift=-0.375em] {RAMPS 1.4};
        
            \draw[arrow] (arduino.20) -- (ramps.west|-arduino.20);
            \draw[arrow] (ramps.west|-arduino.5) -- (arduino.5);

            \draw[arrow] (arduino.20) -| (fpga.105);
            \draw[arrow] (ramps.west|-arduino.5) -| (fpga.75);
        \end{tikzpicture}
        \caption{FPGA for signal recording.}
    \end{subfigure}
    \caption{Different signal path options on the \textsc{OffRAMPS}.}
    \label{fig:signal_block_diagram}
\end{figure}
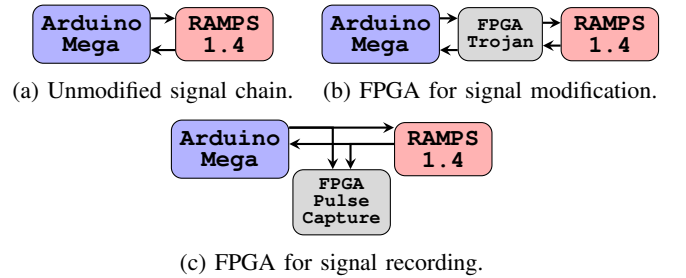

RAMPS 1.4 is an open-source control PCB for 3D printers, designed by RepRap, made to interface with an Arduino Mega by plugging into the top as a HAT device (Figures~\ref{fig:ramps-1}, \ref{fig:ramps-2}).
We make use of this setup for the \textsc{OffRAMPS} as a representation of control boards used throughout the industry\newt{}{, as all FFF printers will ultimately require the same set of signals}.
\textsc{OffRAMPS} allows the Arduino/RAMPS stack to function normally when in a standard configuration, but enables the FPGA to analyze and modify the signals passing between the original boards with simple jumper changes (the yellow highlighted banks in~\Cref{fig:offramps}).
These can redirect signals from normal operation to MITM operation, changing source and destination as needed. 
Meanwhile, the FPGA is used as a reconfigurable platform to enable adversarial and defensive techniques to be tested, analyzed, and verified in hardware.
\Cref{fig:signal_block_diagram} shows the three signal path configurations possible with the added MITM FPGA.

\subsection{Primary \textsc{OffRAMPS} Printed Circuit Board Design}
\textsc{OffRAMPS} has the following major components:
\subsubsection{Digilent Cmod-A7}

The Cmod-A7 FPGA development board~\cite{digilent_cmod_2019} has a Xilinx Artix-7 35T FPGA, LEDs and two push buttons. It is used as the MITM deployment platform between the Arduino and the RAMPS boards.

To intercept all 3D printer control signals between the Arduino and RAMPS boards, all GPIO headers were used, save one pin which was broken out to assist in debugging or external signal insertion.
Additionally, the Artix-7 has an onboard analog-to-digital converter (ADC) which can be used alongside an off-chip digital-to-analog converter (DAC) and opamp to read and modify analog signals like the voltage level through the thermistors on the 3D printer.

\subsubsection{Arduino Mega}
The Arduino Mega~\cite{arduino_mega_2023} in this system is configured to run the popular 3D-printer open-source  Marlin firmware~\cite{marlinfirmware_marlin_2023}.
It receives \texttt{g-code} generated by a slicer such as Cura or Slic3r and sends signals to drive components on the connected 3D printer.
These include (1) step and direction (\texttt{$\ast$\_STEP} and \texttt{$\ast$\_DIR}) signals for each of the stepper motors on the printer---\texttt{X,Y,Z} axis movement and filament extrusion; (2) fan speed control for the part cooling fan, (3) heating element control for both the heated bed and hotend, and (4) UART signals to interact with a connected display/control board.
Marlin implements some safety features such as checks for thermal runaway.

\subsubsection{RAMPS 1.4}
The RAMPS board controls the actuator functions of the printer directly with stepper motor drivers, fan control circuity, and heating element circuitry---all driven by the aforementioned signals sent from the Arduino.
In turn this board sends back signals for the endstops of the axes and the thermistors for both the heated bed and hotend of the printer.
The display/control board also connects through the RAMPS using UART to allow a user to interact with the printer directly without having a connected host computer.

RAMPS has onboard configuration jumpers to micro step the stepper motor drivers, but otherwise all control for the onboard devices is managed by the Arduino.
The stepper motor drivers are also modular, we opted to use the default A4988 drivers shipped with RAMPS.
These are inexpensive and popular, representative of components common to commercial 3D printers.



\subsubsection{Logic level shifting} Both the Arduino and RAMPS 1.4 boards require a $5V$ logic level for their signals. 
The Cmod-A7, however, can only support voltages up to $3.3V$. To accommodate this necessary level shift for the FPGA I/O bidirectional logic level shifters and enhanced field effect transistors (FETs) are integrated on board to allow the $5V$ logic to be shifted to a usable $3.3V$ for the FPGA, and then re-converted back to $5V$ for the Arduino and RAMPS boards.

\subsubsection{Board Power} The RAMPS board receives its power from the 3D printer's $24V$ power supply. 
The Arduino and FPGA need a separate power source, which can come from either the USB ports on each board or can be externally and separately provided. 
By separating the power systems and allowing their source to be selected, the \textsc{OffRAMPS} can function without a host computer---a common deployment strategy for many 3D printers.

\subsection{Test Environment}
The test environment used for validating and experimenting with the \textsc{OffRAMPS} consisted of a modified Prusa i3 MK3S+ 3D printer~\cite{prusa_original_2023} and a host computer running Ubuntu 22.04.
The Prusa i3 MK3S+ is an incredibly popular hobbyist-grade 3D printer compliant with the RepRap project.
A small modification had to be made to the Prusa to add mechanical endstops as this is a more common method of homing than the more advanced sensorless homing that the Prusa control board supports natively.
The RAMPS required small modifications to ensure compatibility with a $24V$ power supply---this was done according to the instructions for this conversion from RepRap.
All prints were sliced with Ultimaker Cura and \texttt{g-code} control was done with Repetier Host.
\section{\textsc{OffRAMPS} for Trojan Insertion}
    \subsection{Objective and Relevance}
    The \textsc{OffRAMPS} presents several advantages over conventional firmware and \texttt{g-code} based Trojans: the FPGA leveraged as the MITM allows for \newt{}{both} fine-grained logic and timing level modification of all control signals at the resolution of the FPGA clock speed ($100MHz$).
    We present several Trojans which take advantage of these benefits and the direct access to fundamental control signals. 

    \subsection{Methodology and Design Considerations}
    The \textsc{OffRAMPS} was evaluated for its ability to implement Trojans mimicking common 3D printer issues as well as Trojans which maliciously compromise the  printer hardware itself, as outlined in~\Cref{tab:Trojans_and_effects}.
    These Trojans are designed to modify the part,  deny access to certain printer elements, or damage the part or the printer itself.
    
    
    \begin{table*}[!ht]
        \centering
        \caption{Trojans evaluated using \textsc{OffRAMPS}. Part modification (PM) Trojans modify the produced part, Denial of Service (DoS) Trojans disable access to a function of the printer, and Destructive (D) Trojans damage a component of the printer itself. Trojans T1 - T5 allow for completed prints which are shown placed on graph paper with line spacing of $\frac{1}{4}$ inch.}
        \label{tab:Trojans_and_effects}
        \begin{subfigure}[t]{\linewidth}
        \SetTblrInner{rowsep=3pt}
        \begin{tblr}{
            colspec = {Q[c, wd=0.05\linewidth]|Q[m, wd=0.035\linewidth]|Q[m, wd=0.07\linewidth]|Q[m, wd=0.2\linewidth]|Q[m, wd=0.41\linewidth] X[m]},
            hlines,
        }
        \toprule
            \textbf{Trojan} & \textbf{Type} & \textbf{Scenario} & \textbf{Effect} & \textbf{Printed Part} \\ \midrule
            
            \textbf{T0} & None & None & Golden print & \parbox[c]{\linewidth}{\frame{\includegraphics[height=4em, width=\linewidth]{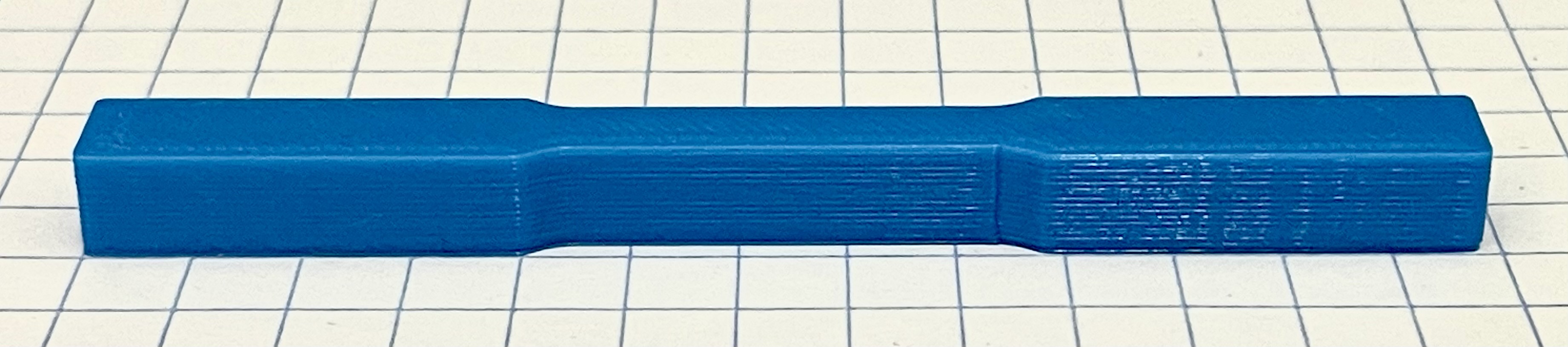}}} & \parbox[c]{\linewidth}{\frame{\includegraphics[height=4em, width=0.65\linewidth]{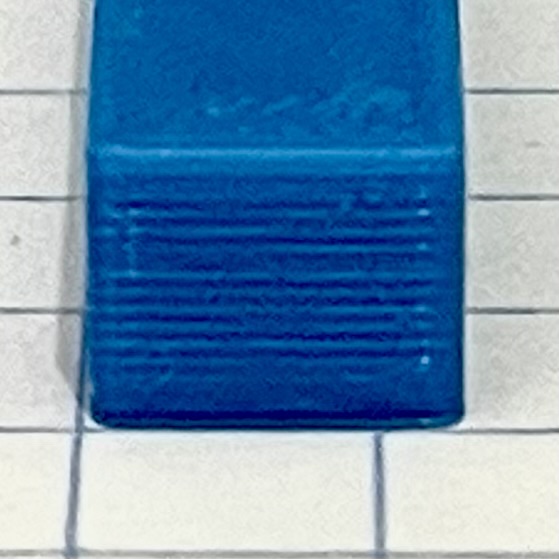}}} \\ \midrule
                                                   
            \textbf{T1} & PM & Loose Belt & Randomly changes steps from X or Y axis during print & \parbox[c]{\linewidth}{\frame{\includegraphics[height=4em, width=\linewidth]{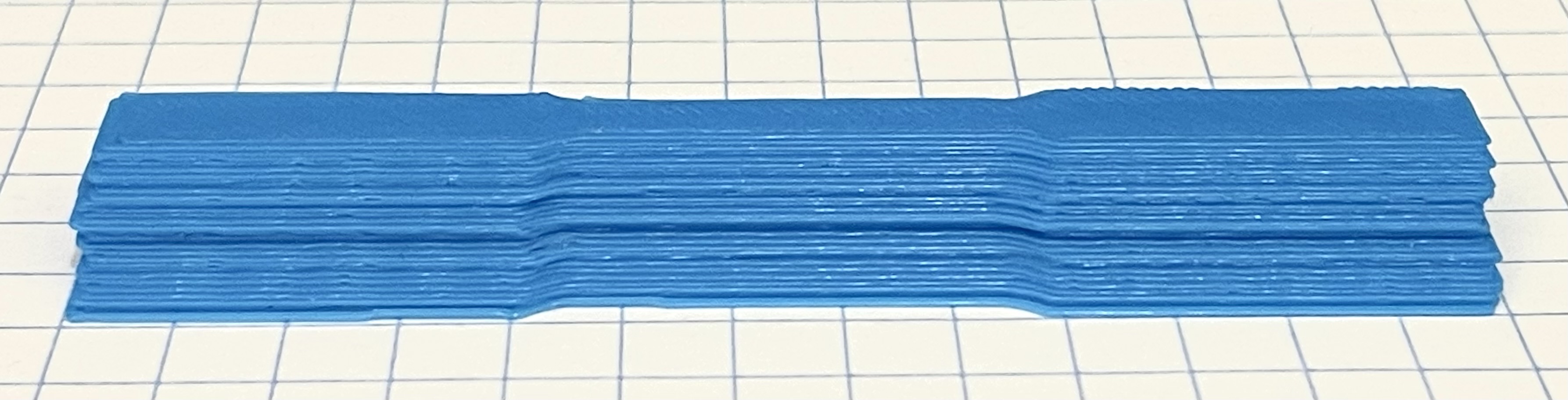}}} &\parbox[c]{\linewidth}{\frame{\includegraphics[height=4em, width=0.65\linewidth]{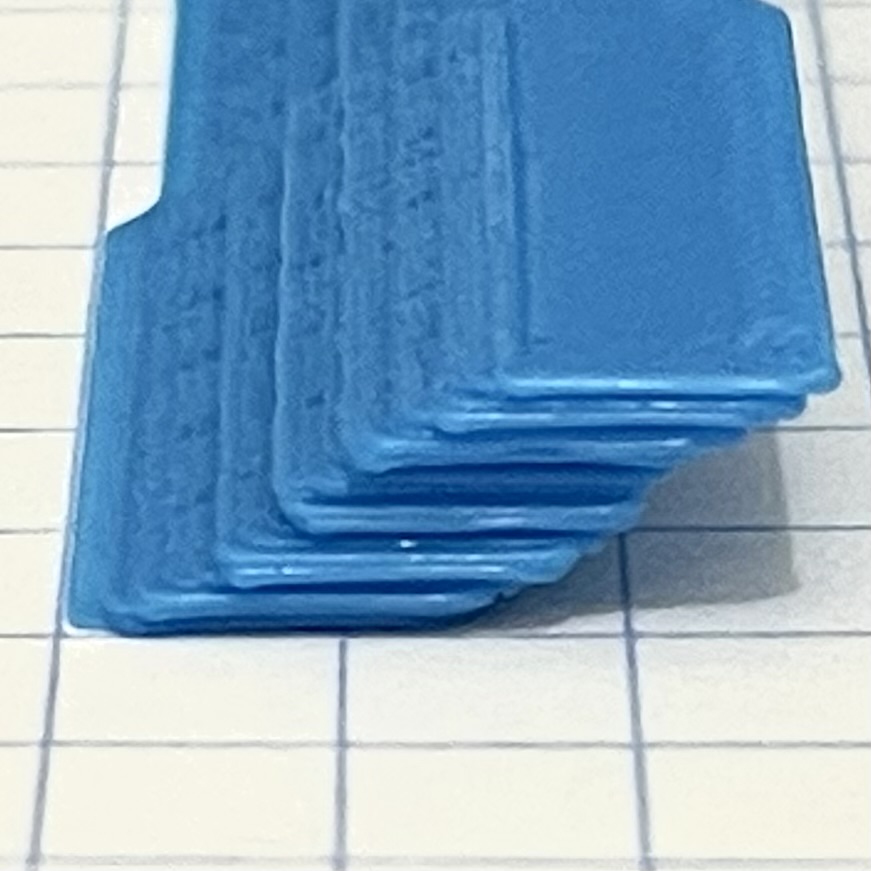}}} \\  \midrule
            \textbf{T2} & PM & Incorrect Slicing & Constant over / under extrusion per print & \parbox[c]{\linewidth}{\frame{\includegraphics[height=4em, width=\linewidth]{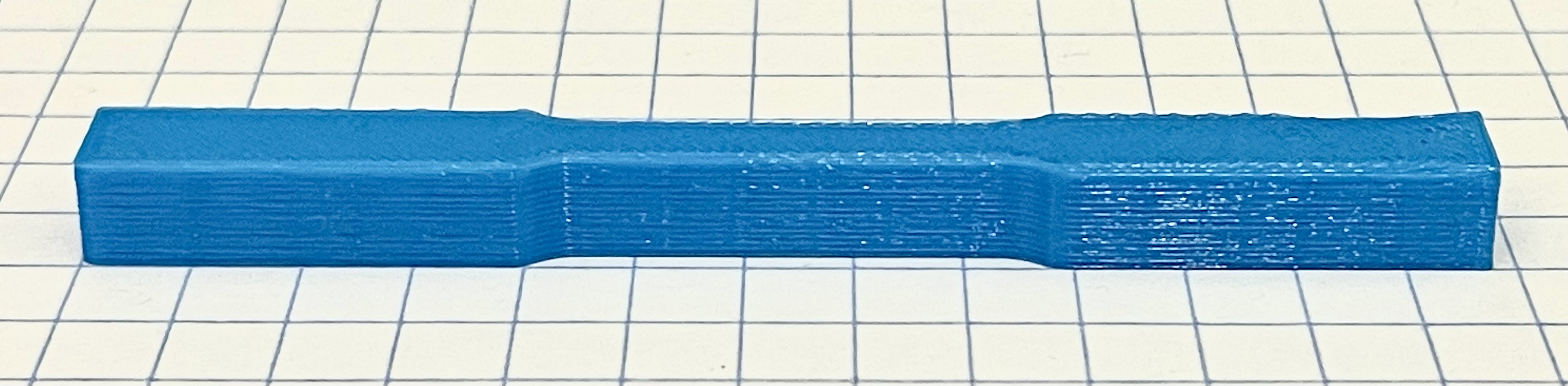}}} &\parbox[c]{\linewidth}{\frame{\includegraphics[height=4em, width=0.65\linewidth]{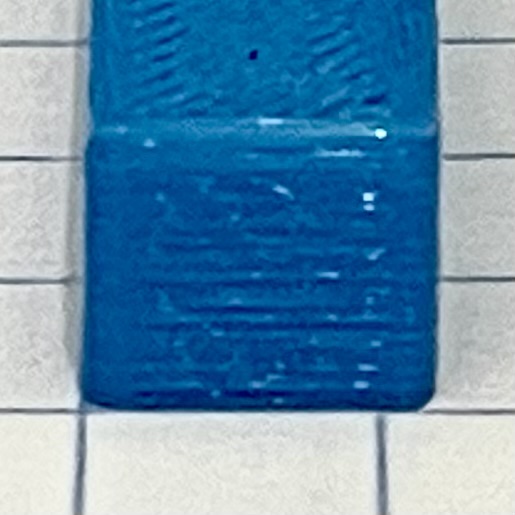}}} \\  \midrule
            \textbf{T3} & PM & Incorrect Slicing & Increases or decreases filament retraction during Y steps & \parbox[c]{\linewidth}{\frame{\includegraphics[height=4em, width=\linewidth]{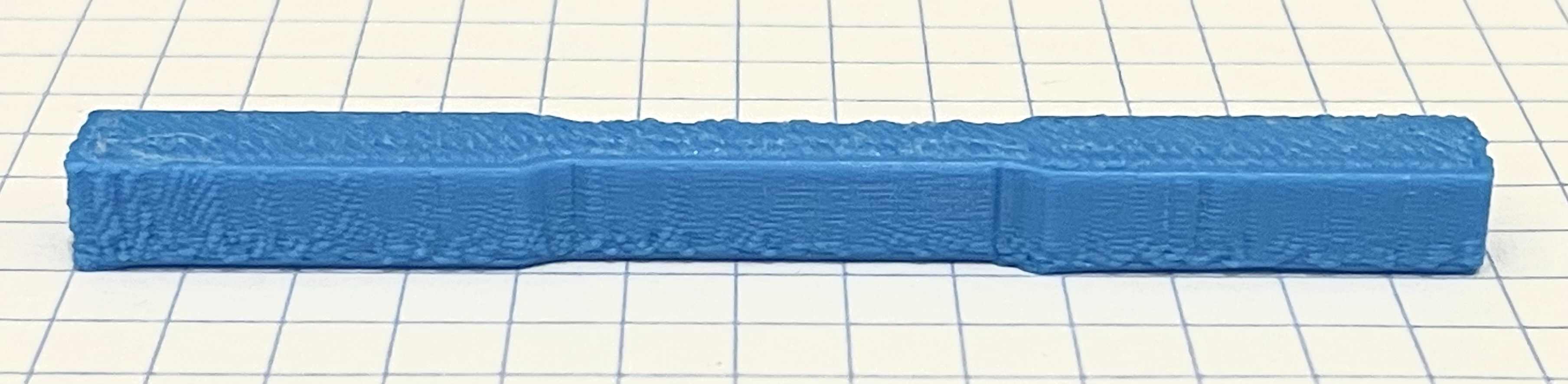}}} &\parbox[c]{\linewidth}{\frame{\includegraphics[height=4em, width=0.65\linewidth]{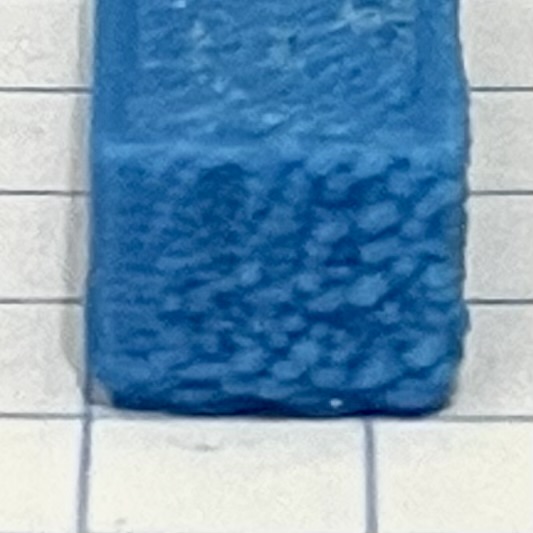}}} \\  \midrule
            \textbf{T4} & PM & Z-Wobble & Small Shift along X and Y axis on random Z layer increments & \parbox[c]{\linewidth}{\frame{\includegraphics[height=4em, width=\linewidth]{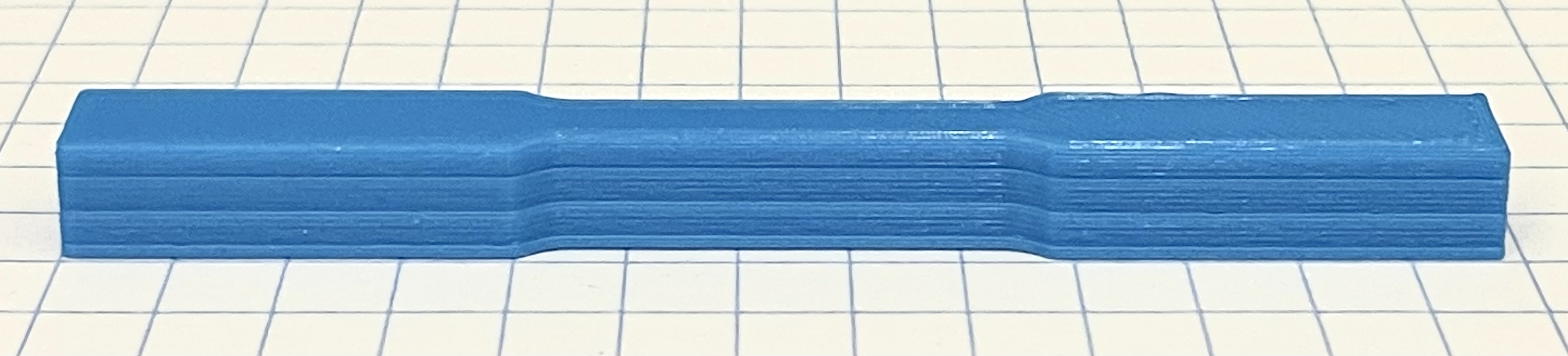}}} &\parbox[c]{\linewidth}{\frame{\includegraphics[height=4em, width=0.65\linewidth]{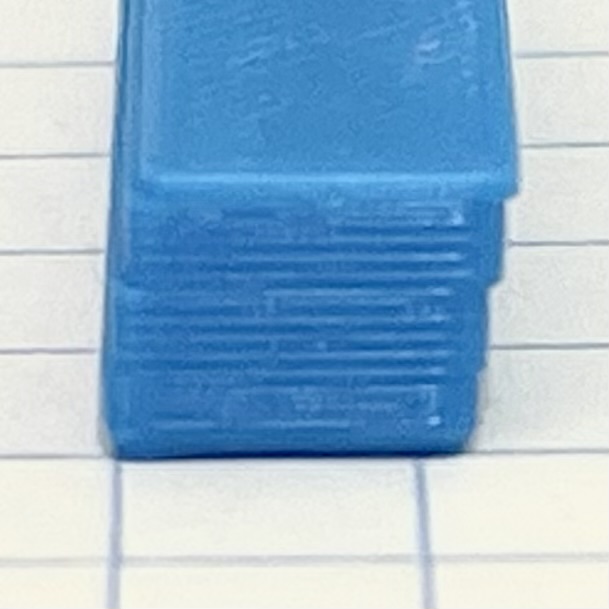}}} \\  \midrule
            \textbf{T5} & PM & Incorrect Slicing & Layer delamination via Z-layer shift & \parbox[c]{\linewidth}{\frame{\includegraphics[height=4em, width=\linewidth]{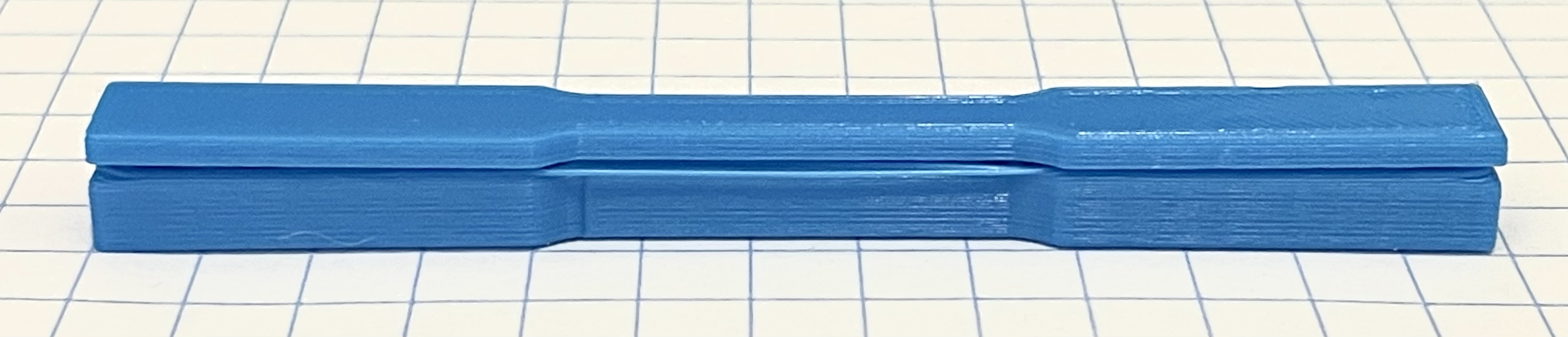}}} &\parbox[c]{\linewidth}{\frame{\includegraphics[height=4em, width=0.65\linewidth]{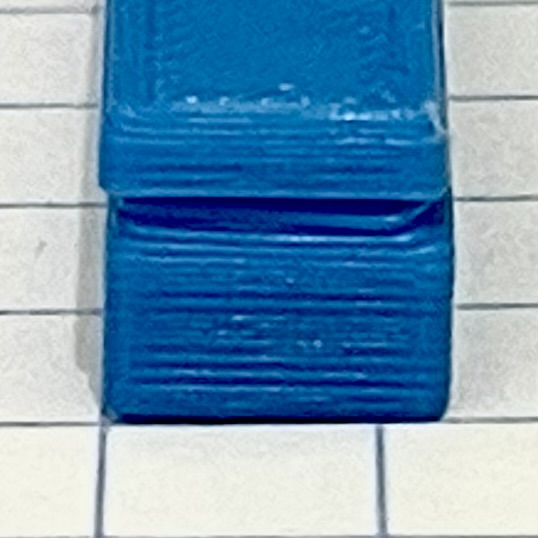}}} \\
        \end{tblr}
         \end{subfigure}

         \begin{subfigure}[t]{\linewidth}
        \begin{tblr}{
            colspec = {Q[c, wd=0.05\linewidth]|Q[m, wd=0.035\linewidth]|Q[m, wd=0.07\linewidth]|Q[m, wd=0.235\linewidth]||Q[c, wd=0.05\linewidth]|Q[m, wd=0.035\linewidth]|Q[m, wd=0.07\linewidth]|X[m]},
            hlines,
        }   
        \midrule
         \textbf{Trojan} & \textbf{Type} & \textbf{Scenario} & \textbf{Effect} & \textbf{Trojan} & \textbf{Type} & \textbf{Scenario} & \textbf{Effect} \\ \midrule
            \textbf{T6} & DoS & Hardware Failure & Denial of service via disabling D8/D10 heating element power & \textbf{T8} & DoS & Hardware Failure & Arbitrarily deactivating stepper motors via EN signals \\  \midrule
            \textbf{T7} & D & Hardware Failure & Forcing thermal runaway and permanently enabling heating elements & \textbf{T9} & PM & Hardware Failure & Arbitrarily reducing part fan speed mid-print \\  \bottomrule
        \end{tblr}
        \end{subfigure}
         
    \end{table*}

 A  framework for the insertion of  Trojans was created using VHDL. This framework allows for both standard operation of the printer via signal bypass or malicious operation through the Trojan module. Several sub-modules were created to control the insertion of Trojans as follows:

    \textbf{Pulse Generation Module} handles the generation of pulses for the stepper motor drivers, and allows for the customization of both frequency and pulse width, along with input parameters for micro stepping determined by the printer configuration.
     
    \textbf{Edge Detection Module} implements an edge detector to identify events such as print head movements or extrusions via observation of the \texttt{STEP} and \texttt{DIR} stepper motor driver signals from the Arduino or endstop actuation for homing detection.

    \textbf{Homing Detection Module} is a state machine which tracks actuation of the endstops in a defined order to determine when the print head has homed. This is the first action taken at the start of print and can determine when to activate Trojans.
    
    \textbf{Trojan Control Module} has logic to enable or disable each of the Trojans (\Cref{tab:Trojans_and_effects}), along with control units for each Trojan. The modified signals produced by this module are multiplexed with the original control signals so the Trojans can be dynamically activated or deactivated.

    \subsection{Results and Analysis of Effectiveness}

\textbf{Golden Print} (\Cref{tab:Trojans_and_effects}: T0), was created by setting the FPGA in `bypass' mode. The control signals from the Arduino pass through, without modification, to the RAMPS board. The printed part shows no deformation, structural compromise, or dimensional inaccuracy.

\textbf{Trojan T1} implements an arbitrary shift along the X and Y axes every ten seconds. This print shows extensive shift along both axes, affecting dimensional accuracy and part practicality. The FPGA on the \textsc{OffRAMPS} allows to injection stepper motor pulses in between the original control pulses, causing longer travel motions of the print head. This effect is used by the Trojan to add extra steps without adding extra print time.

\textbf{Trojan T2} modifies the amount of material extruded during printing, similar to a `flow' parameter used when slicing a STL model. The Trojaned part, was printed while masking half of extruder stepper motor pulses sent to the RAMPS board, reducing the flow and amount of material extruded by 50\%. This implements reduction Trojans from Flaw3D.

\textbf{Trojan T3} mimics a type of problem which can occur from improper settings when slicing a model into \texttt{g-code}. Retraction refers to the amount of filament that is pulled back during certain movements. By affecting extruder steps during some movements we can cause over or under extrusion in a way that could appear to a user as if part settings were incorrect when sliced. This Trojan is shown with over extrusion in \Cref{tab:Trojans_and_effects}: T3.

\textbf{Trojan T4} implements a Z-wobble Trojan. Z-wobble is  common build issue with 3D printers, where the frame holding the Z-axis is not rigid; thus, the print head can shift during printing. Trojan T4 emulates this error by adding steps on one axis during printing causing layer shifts.

\textbf{Trojan T5} causes an arbitrarily sized shift on the Z-axis, causing poor layer adhesion or, in severe cases, layer delamination. This mimics improper slicing settings if the layer spacing is modified throughout the print, and poor hardware setup if a shift is done at the start of print, causing the part to fail to adhere to build plate. 
        
\textbf{Trojan T6:} 3D printers have multiple heating elements including a heated bed, to assist with part adhesion to the build plate, and a hotend for extruding molten material. Should these heating elements be unable to reach the necessary temperature, the 3D printer may be unable to begin a print or if the temperature should suddenly drop mid-print the part quality could be severely negatively impacted. This can also cause the firmware to throw a thermal runaway error and halt all operation. This Trojan was observed to successfully turn off the PID controlled MOSFETs employed in providing power to the heating elements, causing the Marlin firmware to enter an error state and end the print prematurely.
        
\textbf{Trojan T7:} The inverse of Trojan T6, Trojan T7 forces the heated elements to continue heating regardless of the firmware temperature control. By implementing this Trojan in hardware we are not only able to force overheating, but also able to ignore the firmware's thermal runaway panic and continue heating the elements. This is a purely destructive Trojan which can not only severely degrade part quality but can damage or destroy components of the printer itself. This Trojan was observed to successfully enable power MOSFETs for the heating elements permanently, bypassing all thermal control and fail-safes from the firmware, heating the element past the working specification. Furthermore, since the MOSFETs are fully turned on at a 100\% duty cycle, the temperature of the hot-end was observed to rise extremely fast, passing the intended temperature within a few seconds of activation.

\textbf{Trojan T8:} Each stepper motor driver has an input signal \texttt{$\ast$\_EN} which determines if the motor is engaged and able to be moved. By actuating this signal throughout the print we can disable stepper motor movements strategically to fail a print. This emulates issues with stepper motor drivers or the motors as they can be made to arbitrarily cease functioning.
        
\textbf{Trojan T9} affects the part-cooling fan on the printer and causes either over- or under-cooling during printing. Depending on the print material and the stage of printing, the cooling fan runs at various speeds, determined by the \texttt{g-code}. Control signals for this fan are passed through the FPGA for full control. Print quality can be degraded by either over- or under-cooling. It can fail  if excessively cooled at the first layer causing it to pull off the build plate.

\newt{}{Trojans \textbf{T1 - T5} produce prints with visible or structural anomalies and were printed using the \textsc{OffRAMPS} with a Prusa i3 MK3S+ printer, shown in~\Cref{tab:Trojans_and_effects}.
Trojans \textbf{T6 - T9} affect aspects of the 3D printer which either prevent printing a part or would be destructive to the printer hardware; they were validated on our printer but did not produce parts we could show.}

\subsection{Key Takeaways}
Low-level access to all control signals afforded by the \textsc{OffRAMPS} board is a powerful capability to implement many different types of Trojans (the listed Trojans are not exhaustive of all possibilities). The FPGA also allows for the insertion of arbitrary combinations of Trojans along with their triggers.  Accessibility and granular control for interception and modification of signals makes the \textsc{OffRAMPS} board a powerful tool in inserting and testing 3D printer Trojans.
    
\section{\textsc{OffRAMPS} for Print Monitoring}
In addition to signal modification in hardware, \textsc{OffRAMPS} also allows for signal extraction, enabling tests for print verification and Trojan detection.
In a sense, the FPGA can act as a rudimentary `digital logic analyzer' for the control signals passing between the Arduino and RAMPS boards.

\subsection{Objective and Relevance}
Most 3D printer Trojans rely on maliciously altering a design at a stage prior to the signal transfer between the firmware and the control circuitry.
Flaw3D~\cite{pearce2022flaw3d}, for example, modifies a design using a malicious bootloader to edit the \texttt{g-code} as it is sent to the firmware, resulting in prints with reduced structural integrity.
Dr0wned~\cite{belikovetsky_dr0wned_2017} creates modified models prior to their slicing, which ultimately will result in improper \texttt{g-code} being sent to the firmware.
By intercepting the signals after they are decoded by the firmware, we are able to record the real movements of the stepper motors and verify them against a known-good model.

\newt{}{This would be useful, for instance, in safety-critical parts in industrial manufacturing. Typically, randomly-selected parts may undergo destructive testing to validate performance. However, if a small subset of parts are intentionally and maliciously defective, then random testing may not identify them. 
\textsc{OffRAMPS} addresses this through continuous monitoring of prints---all parts will be checked, not just a random subset. Further, parts are checked during production, meaning that large malicious divergences can be detected and aborted early to save machine time and material cost.} 

\subsection{Methodology and Design Considerations}
To monitor the 3D printer in real-time, the FPGA is programmed to both record and export the relevant control information.
We designed a module to  track the number of steps sent to each stepper driver after homing. This translates into absolute positions within the build volume of the printer and a definite amount of extruded material.
\newt{This information is captured for known-good (``golden'') prints and used as a point of comparison against which all other prints would be evaluated.}{
The detection algorithm works as follows: (1) a ``golden'' model is captured by verifying a set of \texttt{g-code}. This is done by first performing a print then completing extensive verification through both non-destructive and destructive testing to ensure the part meets the physical demands and constraints of the product. 
(2) Once assured, the pulse profile can be used as a point of comparison for future prints.}

\newt{To acquire these pulse profiles we created a module to track the absolute position of each motor and use the USB-UART peripheral  on the Cmod-A7 to transmit it.}{}

\textbf{Axis Tracking\newt{}{:}} \newt{}{This} module analyzes the stepper motor control signals, \texttt{STEP} and \texttt{DIR}, for each of the axes  and the extruder to determine their positions.
This consists of a set of rising edge detectors and counters, which increment for each \texttt{STEP} rising edge when \texttt{DIR} dictated that the motors were moving in the positive direction and decrement when they moved negatively.
By correlating the steps to the movement along each axis---information available with printer setups---we are able to track the absolute position of the print head within the build volume and the amount of filament extruded.

To ensure the step counting always began from 0 in a known location, we leveraged the homing detection module created for Trojan insertion.
When the printer is homed at the beginning of each print, the step counts and UART transaction counter are initialized.
As the number of steps to home is determined by the arbitrary position of the print head at the start of the print, capturing this data was deemed unnecessary for evaluating the Trojans.

\textbf{UART:} For accurate  pulse counts between all tests, the counter to determine the frequency of the UART transactions starts after the print head is homed and the first \texttt{STEP} edge is found.
This synchronization  significantly increased accuracy over initial tests which did not wait for the first step before beginning the counter.
With the analysis started, the UART control unit sends a 16-byte transaction containing step counts for all of the motors each 0.1 seconds.

\newt{}{\textbf{Overhead:} A detection method which significantly impacts the speed or quality of a print is counterproductive to the goal of verifying that a print is of good quality. We estimated that the maximum propagation delay of any signal captured in the detection design is $12.923ns$ on the \texttt{Y\_DIR} signal. The ordinary signals between the Arduino and RAMPS boards were measured to have maximum frequencies less than $20kHz$ with a minimum pulse width of $1\mu s$. Given these parameters, a $12.923ns$ delay is negligible and we found no effect on print quality while running our detection hardware.}

\subsection{Detecting Trojan-Induced Edits}
Our Trojan detection strategy compares the captured pulse counts of a given print against a known-good capture, either derived from a print that was captured and then separately validated or from a simulation of the firmware. In a print without Trojan manipulation the Arduino will always send the same quantity of \texttt{STEP} and \texttt{DIR} control signals with approximately the same timing as the known-good print, but where the print commands have been interfered with these counts will change. 
Mismatches outside of a reasonable margin of error \newt{occur point}{}suggest this kind of interference, and in our study, this translates to the presence of a Trojan. 
Here, the margin of error is due to the \newt{previously-discussed}{}challenge of synchronizing the step counting with the UART transactions.
\newt{Some drift in the counts occurs over the course of even known-good test prints.}{Additive manufacturing systems are asynchronous, so an instruction can take a slightly different amount of time when executed multiple times or across multiple prints.
This variation, referred to as ``time noise''~\cite{liang_practical_2021}, means that some drift in the step counts will occur over the course of even known-good test prints.}
This drift was, however, always less than a 5\% difference in our testing, so for our evaluations we used this 5\% margin of error against our ideal profile.
\newt{}{This 5\% margin of error can be made significantly smaller with a faster communication protocol, as fewer steps possible per transaction would lower the potential drift in counts.
The concern of having too large of an error margin is also mitigated with a final check with a 0\% margin of error, ensuring that the correct number of steps was counted on each axis at the conclusion of the print.}

A Python script compares a newly captured print against a ``golden'' model.
Should a mismatch outside of the 5\% margin of error occur the transaction number and mismatching values are printed.
At the termination of the capture file the script then gives a report stating the total number of mismatches, the greatest error found, and the total number of captured transactions---based on these a determination of whether or not a Trojan is suspected is made.
This analysis can also be done in real-time while printing, enabling a user to halt a print as soon as a Trojan is suspected.

\subsection{Analysis of Flaw3D Trojans}
To evaluate the Trojan detection methodology we emulated Trojans from Flaw3D~\cite{pearce2022flaw3d}.
In the original work a modified bootloader was used to change \texttt{g-code} on the fly to implement one of two types of Trojan: reduction of extruded filament or occasional relocation of filament during the print.
We recreate these Trojans using a Python script which modifies given \texttt{g-code} in the same way the malicious bootloader does.
This yielded eight Trojans from two categories, each with varying levels of severity as
enumerated in~\Cref{tab:flaw3d_Trojans}.

\begin{table}[t]
    \centering
    \caption{Flaw3D Trojans. Modification value for reduction is a factor by which extrusion amount is reduced. For relocation it is the number of movements before filament is relocated.}
    \begin{tabular}{c|c|c|c}
        \toprule
         \textbf{Test Case} & \textbf{Type} & \textbf{Modification Value} & \textbf{Detected} \\ \midrule
         1              & Reduction     & 0.5       &   {\color{mygreen}\ding{51}}\\
         2              & Reduction     & 0.85      &   {\color{mygreen}\ding{51}}\\
         3              & Reduction     & 0.9       &   {\color{mygreen}\ding{51}}\\
         4              & Reduction     & 0.98      &   {\color{mygreen}\ding{51}}\\ \midrule
         5              & Relocation    & 5         &   {\color{mygreen}\ding{51}}\\
         6              & Relocation    & 10        &   {\color{mygreen}\ding{51}}\\
         7              & Relocation    & 20        &   {\color{mygreen}\ding{51}}\\
         8              & Relocation    & 100       &   {\color{mygreen}\ding{51}}\\ \bottomrule
    \end{tabular}
    \label{tab:flaw3d_Trojans}
    \vspace{-3mm}
\end{table}

Each of these Trojans was printed and their pulse profiles were captured using the \textsc{OffRAMPS}.
Those captures were then compared against the known-good reference and the detection program was able to identify all of the Trojans. A selection of the captures and tool output is given in~\Cref{fig:Trojan_detected} showing mismatches outside of the margin of error and the detection tool output identifying them.
Here, the Trojan used was Test Case 7 (\Cref{tab:flaw3d_Trojans}), which is not a stealthy Trojan.

The stealthiest Trojans tested are Test cases 4 and 8 which reduce extrusion by only 2\% and relocate material every 100 moves, respectively.
In both, the Trojan is minimal enough that structural integrity was not noticeably impacted.
However, the detection strategy was still able to identify their presence.

\lstdefinestyle{mystyle}{
    moredelim=[is][\textcolor{mygreen}]{!*}{*!},
    moredelim=[is][\textcolor{red}]{@*}{*@},
}
\begin{figure}[b]
    \centering
    \begin{subfigure}{0.43\linewidth}
    \begin{tikzpicture}
            \node[anchor=south west, inner sep=0] (image) at (0,0) {
            \begin{lstlisting}[style=mystyle]
Index, X,    Y,    Z,   E
5113,  6060, 8266, 960, 52843
5114,  6304, 8095, 960, 52856
5115,  !*7218*!, 8285, 960, 52856
5116,  !*8166*!, 8483, 960, 52856
5117,  8671, 8620, 960, 52859
5118,  8384, 8733, 960, 52875
        \end{lstlisting}      
            };           
            \begin{scope}[overlay]
                \fill[mygreen] ([xshift=-2.6mm,yshift=8.88mm] image.south west) -- +(0,2.2mm) -- +(1.45mm,1.1mm) -- cycle;
                \fill[mygreen] ([xshift=-2.6mm,yshift=6.52mm] image.south west) -- +(0,2.2mm) -- +(1.45mm,1.1mm) -- cycle;
                \draw[mygreen, line width=1pt](0.85,0.6) rectangle (1.55,1.1);
            \end{scope}
        \end{tikzpicture}
        \caption{Selection of transactions from the golden reference.}
        \label{fig:known_ref}
        \vspace{2mm}
    \end{subfigure}
    \hspace{1mm}
    \begin{subfigure}{0.43\linewidth}
        \begin{tikzpicture}
            \node[anchor=south west, inner sep=0] (image) at (0,0) {
            \begin{lstlisting}[style=mystyle]
Index, X,    Y,    Z,   E
5113,  6027, 8499, 960, 52832
5114,  6113, 8213, 960, 52846
5115,  @*6489*@, 8133, 960, 52856
5116,  @*7437*@, 8331, 960, 52856
5117,  8384, 8528, 960, 52856
5118,  8601, 8644, 960, 52863
        \end{lstlisting}      
            };           
            \begin{scope}[overlay]
                \fill[red] ([xshift=-2.6mm,yshift=8.88mm] image.south west) -- +(0,2.2mm) -- +(1.45mm,1.1mm) -- cycle;
                \fill[red] ([xshift=-2.6mm,yshift=6.52mm] image.south west) -- +(0,2.2mm) -- +(1.45mm,1.1mm) -- cycle;
                \draw[red, line width=1pt](0.85,0.6) rectangle (1.55,1.1);
            \end{scope}
        \end{tikzpicture}
        \caption{Selection of transactions from Flaw3D Trojan print.}
        \label{fig:troj_capture}
        \vspace{2mm}
    \end{subfigure}
    \begin{subfigure}{0.9\linewidth}
        \begin{tikzpicture}
            \node[anchor=south west, inner sep=0] (image) at (0,0) {
            \begin{lstlisting}[style=mystyle]
...
@*Index: 5115, Column: X, Values: 7218, 6489*@
@*Index: 5116, Column: X, Values: 8166, 7437*@
...
Largest percent difference found: 93.19%
Number of transactions compared: 12416
Number of mismatches: 952
Trojan likely!
        \end{lstlisting}      
            };           
        \end{tikzpicture}

        \caption{Selected output of the Trojan detection tool identifying a mismatch of transactions on the X axis with index 5115 and 5116. These values fall outside of the 5\% margin of error, so the tool reports a Trojan is likely alongside other metadata. } 
    \end{subfigure}
    \caption{Detection of an emulated Flaw3D Trojan which relocates material every 20 movements.}
    \label{fig:Trojan_detected}
    \vspace{-1mm}
\end{figure}

\newt{}{We did not use this detection method to evaluate our own Trojans, as both the attacks and defense would be co-located in the same FPGA and we do not believe this would demonstrate any meaningful capabilities.}
 
\section{Discussion}

\textbf{Evaluation:} The \textsc{OffRAMPS} was successfully used to both insert and detect 3D printer Trojans and, with the reconfigurability of the MITM FPGA, could implement more novel Trojans, requiring fine-grained manipulation and analysis of the firmware-produced control signals.
This platform provides a basis for considerable future experimentation, with expansion of both the kinds of attacks which may be undertaken as well as new golden-free methods for detection and even reverse-engineering printed parts from their control signals.

Given the increasing adoption of Additive Manufacturing for safety-critical components and commercial applications, facilitating the analysis of security vulnerabilities in printers and real-time validation of part printing enables designs to be produced safely by providing methods to detect interference.

\textbf{Limitations:} While \textsc{OffRAMPS} was able to detect the tested firmware Trojans from literature, the study still has some design limitations. Firstly, the platform is limited in its ability to relay high-speed information (i.e. a high-frequency data capture) to a host PC due to a lack of circuitry to support a high speed communication interface such as Ethernet or USB, preventing complex analysis strategies.
Though it can \textit{emulate} them, \textsc{OffRAMPS} is currently unable to \textit{detect} any Trojans which affect the heating elements, whether implemented in firmware or hardware.
In addition, though the platform is designed with power isolation between the major components and can also support undervolting and brown-out attacks, this study did not explore this area, nor is there suitable circuitry for detecting such an attack.
Many 3D printers are intended to be run while not actively connected to a host computer, which the \textsc{OffRAMPS} cannot currently support for its Trojan detection functionality.

\newt{}{\textbf{Related platforms:} Other methods of attacks and defenses exist but are predominantly based on lossy side-channels such as acoustic, power, electromagnetic emission, or optical analysis.
The \textsc{OffRAMPS}, by connecting directly to control signals, is uniquely able to modify or analyze prints with no loss of data.
Including support for some of these other side-channel techniques is being considered for future revisions, but we have found no examples of other hardware platforms which can be used for 3D printer attacking or modifying in the same manner as the \textsc{OffRAMPS}.}

\textbf{Responsible disclosure:} As all studied attacks required modification to the underlying components, no responsible disclosure is necessary for this work.


\section{Conclusion}
In this work we present \textsc{OffRAMPS}, a new hardware tool for emulation, evaluation, and detection of 3D printer Trojans.
By using an FPGA in a machine-in-the-middle configuration we are able to dynamically modify 3D printer control signals post-firmware as well as detect Trojans implemented at or before the firmware level. This enables investigation of both attack and defense scenarios, a task otherwise complicated by the relationship between digital and real-world components.

Using \textsc{OffRAMPS} we are able to emulate existing Trojans from the literature in hardware, as well as implement new Trojans which cannot easily be done in firmware. In total we implemented 9 such attacks from simple denial-of-service to subtle part modifications and thermal runaway, the largest suite ever supported by a single platform.
On the defensive side, we present a simple yet effective strategy which can be implemented in \textsc{OffRAMPS} to count the number of pulses over a series of time windows and compare this against a golden series of data (which can come from simulation), this strategy could detect all Trojans from the literature.


\section*{Acknowledgements}
This work was supported in part by DoE Kansas City. Honeywell Federal Manufacturing \& Technologies, LLC operates the Kansas City National Security Campus for the United States Department of Energy / National Nuclear Security Administration under Contract Number DE-NA0002839.

\bibliographystyle{IEEEtran}
\bibliography{bib/benhamram,bib/printing_ref,bib/am-firmware-attacks}

\end{document}